# Dirac surface states and nature of superconductivity in Noncentrosymmetric BiPd

Zhixiang Sun[1], Mostafa Enayat[1], Ana Maldonado[1,2], Calum Lithgow[2,3], Ed Yelland[2], Darren C. Peets[1,4], Alexander Yaresko[1], Andreas P. Schnyder[1] & Peter Wahl[1,5]

In non-magnetic bulk materials, inversion symmetry protects the spin degeneracy. If the bulk crystal structure lacks a centre of inversion, however, spin–orbit interactions lift the spin degeneracy, leading to a Rashba metal whose Fermi surfaces exhibit an intricate spin texture. In superconducting Rashba metals a pairing wavefunction constructed from these complex spin structures will generally contain both singlet and triplet character. Here we examine the possible triplet components of the order parameter in noncentrosymmetric BiPd, combining for the first time in a noncentrosymmetric superconductor macroscopic characterization, atomic-scale ultra-low-temperature scanning tunnelling spectroscopy, and relativistic first-principles calculations. While the superconducting state of BiPd appears topologically trivial, consistent with Bardeen–Cooper–Schrieffer theory with an order parameter governed by a single isotropic s-wave gap, we show that the material exhibits Dirac-cone surface states with a helical spin polarization.

[1] Max-Planck-Institut für Festkörperforschung, Heisenbergstrasse 1, D-70569 Stuttgart, Germany. [2] School of Physics and Astronomy, University of St Andrews, North Haugh, St Andrews, Fife KY16 9SS, UK. [3] School of Physics and Astronomy, University of Edinburgh, Mayfield Road, Edinburgh, EH9 3JZ UK. [4] Center for Correlated Electron Systems, Institute for Basic Science, Seoul National University, Seoul 151-747 Korea. [5] SUPA, School of Physics and Astronomy, University of St Andrews, North Haugh, St Andrews, Fife KY16 9SS, UK. Correspondence and requests for materials should be addressed to A.P.S. (email: a.schnyder@fkf.mpg.de) or to P.W. (email: wahl@st-andrews.ac.uk).

In noncentrosymmetric superconductors, the spin–orbit interactions can give rise to a mixing of spin-singlet and spin-triplet pairing components[1–3] and to a topologically nontrivial superconducting phase[4,5], which is characterized by protected Majorana surface states[5–9]. While in several materials indications for a triplet component of the order parameter have been detected, unambiguous evidence remains scarce. Scanning tunnelling microscopy (STM) and spectroscopy (STS) can help to identify the symmetry of the order parameters[10], however, it requires high-quality surfaces, which have not previously been obtained for any noncentrosymmetric superconductor.

The noncentrosymmetric compound BiPd[11–13], which becomes superconducting below $T_c \simeq 3.8$ K (refs 14–17), offers the opportunity to study the interplay of the aforementioned spin–orbit coupling (SOC) effects with superconductivity. The large atomic spin–orbit interaction of the heavy element Bi results in a sizeable Rashba-like spin splitting of the calculated bulk bands and, moreover, may give rise to nontrivial band topologies and unconventional superconducting pairing symmetries. Here the term Rashba is used very broadly to denote any antisymmetric SOC that leads to band dispersions with nontrivial spin textures and a dominant linear term at time-reversal invariant momenta of the Brillouin zone (BZ, a classification of spin–orbit interactions in terms of point-group symmetries is discussed in ref. 18). This may give rise to nontrivial band topologies and unconventional superconducting pairing symmetries. Recent point-contact Andreev-reflection measurements have detected zero-bias anomalies and signs of multigap superconductivity[16]. The former have been interpreted as evidence for Majorana midgap states predicted in topological superconductors.

In the following, we study the interplay of spin–orbit interactions with superconductivity in BiPd using scanning tunnelling spectroscopy and relativistic first-principles calculations. Tunnelling spectroscopy, performed on atomically flat surfaces, shows that superconductivity in BiPd is entirely consistent with conventional s-wave pairing. While the superconducting state of BiPd is topologically trivial, our electronic structure calculations show that the large atomic SOC of Bi alters the normal-state band topology of BiPd, leading to the appearance of a Dirac-cone surface state. The existence of these Dirac surface states is confirmed by our scanning tunnelling measurements of the surface density of states.

## Results

**Topographic images and surface terminations.** α-BiPd has a noncentrosymmetric monoclinic crystal structure with space group P2$_1$ (refs 11,12,13). As shown in Fig. 1a the unit cell contains 16 atoms with four inequivalent sites for each element. The structure is characterized by two double layers stacked along the $b$ axis, related by a 180° screw symmetry. The crystal cleaves preferentially along (010) planes between the Bi layers; due to the lack of inversion symmetry, the (010) and (0$\bar{1}$0) surfaces (facing up and down, respectively, in Fig. 1a) are not equivalent. Topographic STM images obtained on the samples cleaved in cryogenic vacuum show atomically flat terraces up to hundreds of nanometres in lateral size separated by steps (Fig. 1b). Line cuts show a step size of ∼5.3 Å between terraces (Fig. 1c), corresponding to half the height of the unit cell along $b$[13]. In atomic resolution images, we observe two different kinds of surfaces. Figure 1d shows a twin boundary between the two. By comparing the surface structures and the atomic corrugations (Fig. 1e) with the expected Bi terminations, we deduce that the surfaces shown in Fig. 1d correspond to (010) on the left and (0$\bar{1}$0) on the right. The main structural difference between the two terminations is the displacement of the atoms in the surface plane: the (0$\bar{1}$0) termination exhibits predominantly lateral displacement of Bi atoms at the surface, whereas for the (010) terminated surface, there is little lateral distortion but substantial vertical buckling of the neighbouring Bi atoms (Supplementary Fig. 1). The atomic resolution indicates high-quality surfaces free from reconstruction, making this the first noncentrosymmetric superconductor suitable for single-particle spectroscopies.

**Surface states with Rashba-like spin splitting.** The spatial and energy dependence of the local density of states was systematically measured using differential conductance (d$I$/d$V$) spectroscopy. While the spectra do not vary significantly as a function of STM tip position, they display pronounced features in their energy dependence. A typical differential conductance measurement covering a large energy range recorded at $T \sim 1.5$ K is shown in Fig. 2a,b. The spectra exhibit a sharp peak at $\sim 0.4$ V, which, by comparison with band structure calculations (Fig. 2c,d), we associate with a van Hove singularity due to the dispersion of a Dirac surface state near the $\bar{\Gamma}$ point of the surface BZ. For a parabolic two-dimensional band with Rashba-like spin splitting, the width of the peak in the density of states, here 24 mV, is an estimate for the spin–orbit splitting[19]. In Fig. 2c,d, we show the projected bulk bands together with the dispersion of the surface states near $E_F$ of (010) and (0$\bar{1}$0) terminated surfaces obtained from a slab calculation along high-symmetry lines of the surface BZ of BiPd. A Dirac-cone surface state appears at the $\bar{\Gamma}$ point within the projected band gap opened by SOC (see Supplementary Figs 2 and 3). This suggests that spin–orbit interactions change the band topology near the time-reversal invariant momentum $\Gamma$ from trivial to nontrivial. The lower branch of the Dirac state is almost completely buried inside the four bulk $p_{1/2}$ bands and the Dirac point, which is protected by time-reversal symmetry, lies just above the top of these bands. Due to the absence of inversion symmetry, the shape of the Dirac states at the top and bottom sides of (010) BiPd are inequivalent. That is, the Dirac state at the (0$\bar{1}$0) surface is shifted downwards by $\sim 75$ meV as compared with that at the (010)-terminated surface. Similarly the spin textures of the Dirac states at opposite sides are expected to be dissimilar and not simply related by symmetry (Fig. 2c,d). As a result, the Dirac electrons of these surface states fractionalize into two inequivalent halves at the top and bottom sides of BiPd.

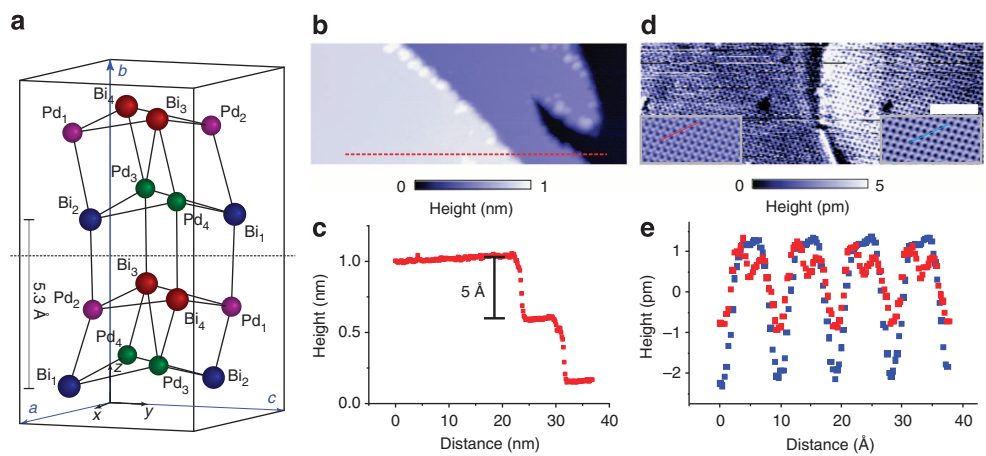

**Figure 1 | Crystal structure and topographic imaging.** (**a**) Crystal structure of BiPd. The dashed horizontal line indicates the preferred cleavage plane between layers of Bi atoms, equivalent planes are at the top and the bottom of the unit cell. Because BiPd is not centrosymmetric, the bottom and top surfaces of a cleaved crystal are not identical. (**b**) STM topography showing atomically flat terraces with step edges, (**c**) height profile extracted at the position indicated by a red dashed line in **b**, showing step edges with ≈5.3 Å height. The step height corresponds to half the height of the unit cell. (**d**) Topographic image of a twin boundary between the two different terminations of the crystal, showing atomic resolution on both of them. The insets show unit-cell averaged close-ups of the atomic corrugation on the two sides—both recorded with the same tip (scale bar, 5 nm). (**e**) Line cut along the lines indicated in panel **d**. From the two line cuts, it can be noted that the difference in height of the two Bi atoms is larger for the surface to the left of the boundary, whereas the modulation of the distance between the Bi atoms is larger for the corrugation on the right. This can be compared with the modulation of the height and interatomic distances expected from the crystal structure and yields an assignment of the left surface to a (010) termination and for the right surface to a (0$\bar{1}$0) termination.

**Properties of the superconducting state.** The next question is whether the Rashba-like spin–split band structure of BiPd manifests exotic superconductivity when cooled. We have characterized the superconducting state in BiPd by ultra-low-temperature STM and STS, transport, specific heat and magnetic susceptibility. Resistivity and magnetic susceptibility (Supplementary Fig. 4a, b), which are in good agreement with previously published results[15–17,20], as well as specific heat (Fig. 3a) show that BiPd becomes superconducting below $T_c \simeq 3.8$ K (Supplementary Note 2). The specific heat data reveal a slightly larger discontinuity at $T_c$ than expected from Bardeen–Cooper–Schrieffer (BCS) theory, while the low-temperature behaviour indicates a full, nodeless gap. Most importantly, the measurements reveal a significant discrepancy between the upper critical field $H_{c2}$ determined from specific heat and transport measurements, as can be seen in Fig. 3a,b. While the magnetic field-dependent resistivity at 1.4 K (Fig. 3b) shows three distinct transitions at 0.1, 0.17 and 0.7 T, the temperature dependence of the specific heat (Fig. 3a) shows no indication of bulk superconductivity at magnetic fields of 0.07 or 0.1 T. This indicates that the transitions observed in the resistance at 0.17 and 0.7 T do not represent bulk behaviour. We note that in ref. 15 the upper critical field based on transport measurements alone was reported as 0.7 T, suggesting a similar non-bulk contribution to the electrical transport may have been present also in those samples. To elucidate the nature of the superconducting order parameter and confirm this result, we have performed STS at temperatures down to 15 mK.

High-resolution tunnelling spectroscopy within a narrow energy range of ±1.25 mV around $E_F$ as shown in Fig. 3c displays a fully-gapped superconducting density of states with sharp coherence peaks. We have fitted a Dynes equation for a single isotropic gap to the tunnelling spectrum, using an effective temperature $T_{eff} = 130$ mK (as determined previously to be the energy resolution of the experiment[21]) and an adjustable broadening parameter $\Gamma$ accounting for impurity scattering, finite quasiparticle lifetime and a possible anisotropy of the order parameter. The fit to the spectrum in Fig. 3c yields a gap with amplitude $\Delta = 601$ µeV and $\Gamma = 12$ µeV. The tunnelling spectra and hence the gap size exhibit very little spatial variation, even across step edges. The temperature dependence of the superconducting gap shows the disappearance of the gap at temperatures close to the bulk transition temperature $T_c$, with temperature dependence in excellent agreement with the BCS theory (Fig. 3d). As opposed to point-contact spectroscopy[16], our data show no evidence for zero-bias peaks when measuring with non-superconducting tips.

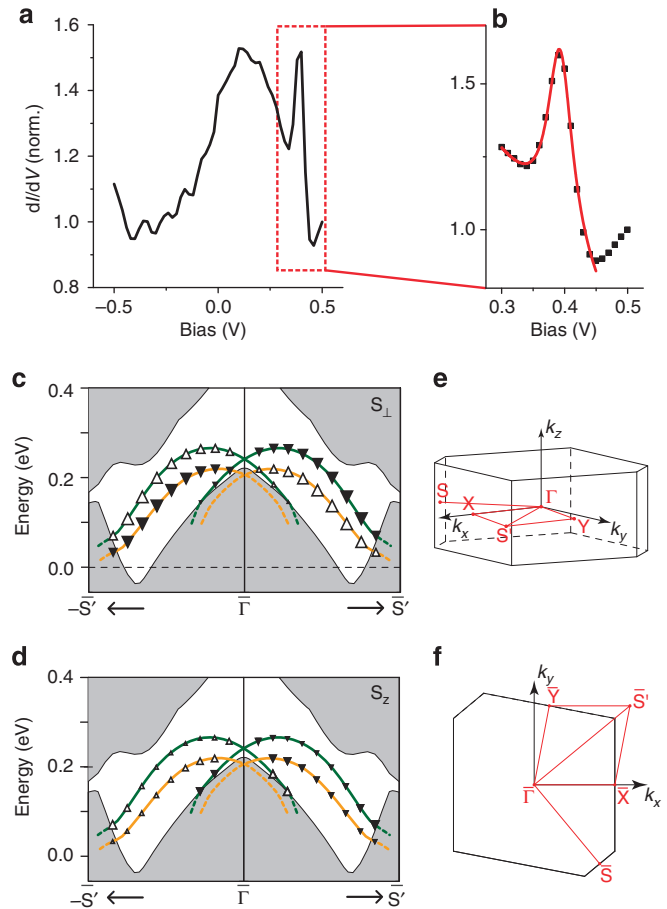

**Figure 2 | Dirac surface states.** (**a**) Tunnelling spectrum obtained on the surface of BiPd, showing a sharp peak at +0.4 V, from atomic resolution images were assigned to a termination as on the left in Fig. 1d. (**b**) High-resolution spectrum around the peak energy that we attribute to a van Hove singularity of the spin–split surface states; a fit of a Lorentzian (red line) to the peak yields a width of 24 mV. Spectra in (**a** and **b**) were taken at a temperature $T \sim 1.5$ K (the resolution of the spectrum in **a** is not sufficiently high to resolve the superconducting gap at 0 V). (**c,d**) Electronic band dispersions near $E_F$ are shown for a (010) slab geometry along the high-symmetry lines of the surface BZ. The bulk bands are shaded in grey and surface states found on the (010) and (0$\bar{1}$0) surfaces as green and orange solid lines, respectively. The spin polarizations of the surface states are indicated by the solid and open triangles. (**c**) The in-plane polarization $S_\perp$ (in the direction perpendicular to $k$) and (**d**) the out-of-plane polarization $S_z$. The spin direction is correlated with the direction of the momentum, forming almost a 90° angle. (**e**) BZ of BiPd for the crystal structure shown in Fig. 1a. (**f**) Surface BZ for the (0$\bar{1}$0) surface.

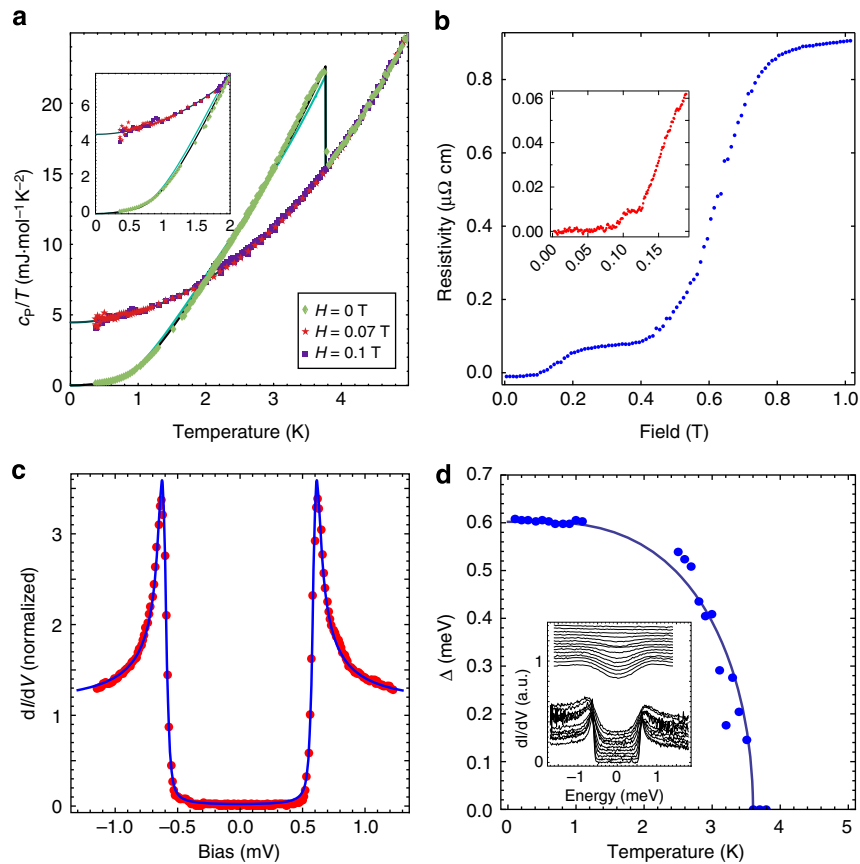

**Figure 3 | Superconductivity.** (**a**) Temperature-dependent specific heat for BiPd, the data show the bulk $T_c$ of 3.8 K, no transition is seen at magnetic fields of 70 mT or larger at temperatures down to 400 mK. Only the lowest critical field seen in transport is consistent with the specific heat. Solid green line shows BCS behaviour, solid black line is calculated using the gap obtained from the tunnelling spectra. (**b**) Transport measurements as a function of magnetic field $B$ performed on the same crystals at 1.4 K. Three transitions are visible when meased at a current $I = 500\,\mu A$ at $H = 0.1$, 0.17 and 0.6 T. At higher currents, the two upper transitions disappear, leaving only the lowest one. (**c**) Tunnelling spectrum of the superconducting state at $T = 15$ mK showing a full gap consistent with isotropic s-wave pairing. The blue line shows the fit of a Dynes function ($I_s = 2.0$ nA, $V_s = 8$ mV and $V_{mod} = 10\,\mu V$). (**d**) Temperature dependence of the superconducting gap, compared with a fit of mean field theory, inset shows the raw spectra.

Tunnelling spectra (Fig. 4a) at 10 mK reveal a complete suppression of the superconducting gap in fields of only 70 mT and indicate an upper limit of the lower critical field $H_{c1}$ of about 20 mT. The field dependence of the gap size, as shown in Fig 4b, confirms that the gap vanishes at an upper critical field $H_{c2} \simeq 75$ mT (extrapolated to $T = 0$ K), which is consistent with the magnetization measurements (compare Supplementary Note 2 and ref. 20) and specific heat. From the tunnelling spectra we extract a ratio of gap size to critical temperature of $\Delta_0/(k_B T_c) \approx 1.9$, slightly larger than the weak coupling BCS prediction of 1.764. Recalculating the behaviour of the specific heat with the experimental gap magnitude results in significantly improved agreement with experiment, confirming that specific heat detects the same superconducting gap size as tunnelling spectroscopy (Fig. 3a, see also Supplementary Note 3 and Supplementary Fig. 5). The level of quantitative agreement among these techniques confirms that tunnelling in BiPd is sensitive to the bulk superconducting properties.

**Imaging of vortex cores and vortex bound states.** In type-II superconductors with nontrivial normal-state band topology, Majorana zero-energy modes can emerge in vortex cores even in the absence of unconventional pairing symmetries[22,23]. To test this possibility, we performed extensive field-dependent STM measurements on the vortex state of BiPd. Figure 4c shows a spatial map of the differential conductance at zero bias, revealing a vortex pinned at a step edge. At the centre of the vortex core we detect a strong zero-bias peak in the differential conductance (Fig. 4d), which we attribute to Caroli-de Gennes–Matricon (CdGM) bound states, of which zero-energy Majorana states are a special case[24–26]. These vortex bound states decay outside the vortex core on the scale of the coherence length $\xi_0$. From the radial dependence of the spectra (see Fig. 4e), $\xi_0$ is estimated to be $\sim 60$ nm, in good agreement with the value obtained from the Ginzburg–Landau expression $\xi_0 = \sqrt{\Phi_0/(2\pi H_{c2})} \simeq 66$ nm ($\Phi_0$ being the magnetic flux quantum $h/(2e)$), reconfirming the extrapolated $H_{c2}$. To distinguish possible zero-energy Majorana modes from ordinary CdGM states with finite energy, a resolution on the order of the level spacing $\varepsilon_0 = \Delta^2/E_F$ of the CdGM states is required. For BiPd, we estimate $\varepsilon_0 \simeq 100$ neV, which cannot be resolved due to limitations of the energy resolution of the instrument, but will also likely be exceeded by the intrinsic broadening of the quasiparticle states. Observation of Majorana modes may require a superconductor with a larger pairing gap.

## Discussion
Our tunnelling spectra, their temperature and magnetic field dependence together with the specific heat data strongly suggest that superconductivity in BiPd behaves rather conventionally, consistent with BCS predictions for a nearly isotropic s-wave order parameter and dominant spin-singlet pairing. A sizeable triplet pairing component, which was recently discussed in the context of topological superconductivity[16,27], is expected to lead to zero-bias anomalies due to Majorana surface states. From our data there is no evidence of a finite surface density of states at zero energy in zero field. While the transition seen in transport at fields $>0.5$ T (Fig. 3b) is consistent with a previously reported critical field of $H_{c2}\approx0.6$ T (ref. 15), STM data and especially specific heat are not. Both yield a substantially smaller critical field $H_{c2}\sim75$ mT, which is also consistent with the vortex core size. This suggests that transport at magnetic fields larger than 75mT and small currents is governed by superconductivity near defects. If the anomalies seen in transport were due to surface effects as might be expected for topological superconductivity, they should leave strong signatures in tunnelling spectra. Point defects or impurities appear unlikely candidates, because a rather large density would be required to dominate the transport. Therefore the more likely explanation is that superconductivity near extended defects, for example twin boundaries, is more robust against magnetic field than in the bulk. In fact, our resistivity measurements come into agreement with other techniques if high-drive currents are used (Supplementary Fig. 4), suggestive of the suppression of critical currents along filamentary pathways. Enhanced critical fields near twin boundaries in noncentrosymmetric superconductors have recently been posited to occur due to magnetoelectric effects[28], while the formation of orbital currents would be suppressed by scattering at the boundaries or by the limited spatial extent of the superconductivity nucleated at such a defect.

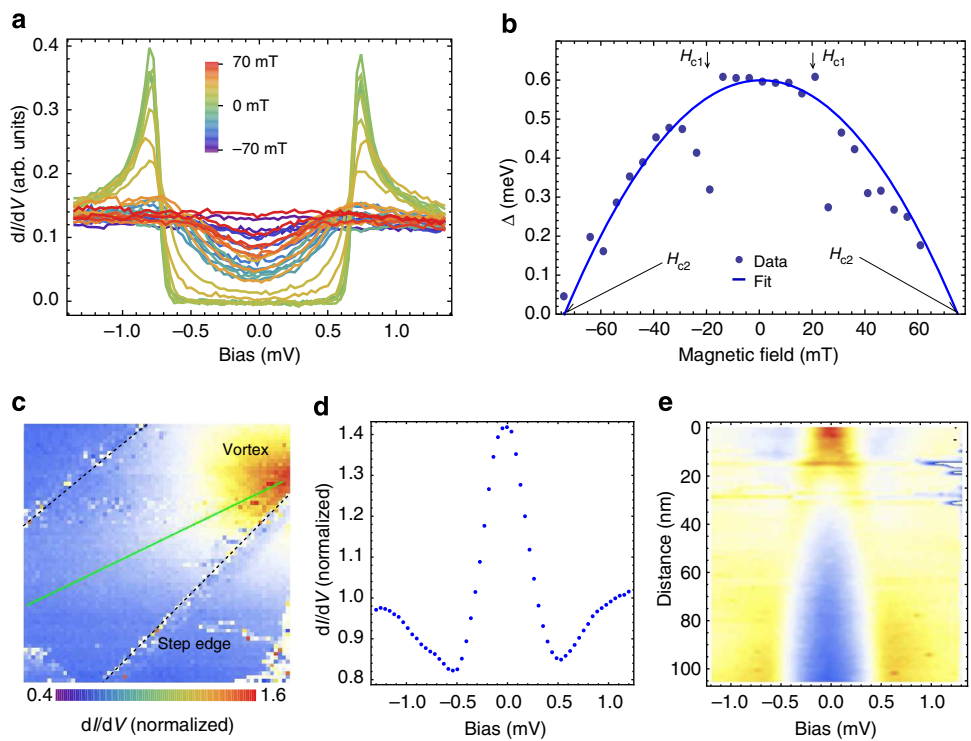

**Figure 4 | Vortex core imaging.** (**a**) STM conductance spectra at base temperature $T=15$ mK as a function of the magnetic field applied perpendicular to the surface ($I_s=2$ nA, $V_s=8$ mV and $V_{mod}=10$ μV). (**b**) Field dependence of the superconducting gap as determined from a BCS fit of the spectra in panel **a**. The blue line represent a fit of $\Delta(H)=\Delta(0)[1-(H/H_{c2})^2]$ to the data points, yielding a critical field $H_{c2}=75$ mT. Arrows mark positions were the data substantially deviate from the expected behaviour, indicating that vortex cores pass by close to the tip position, the lowest fields at which these are detected yield an upper limit for $H_{c1}<20$ mT. (**c**) Differential conductance map at the Fermi energy with an applied magnetic field of $H=53$ mT along the b axis ($V_B=0$ mV, $I_s=2$ nA, $V_s=8$ mV and $V_{mod}=50$ μV, $100\times100$ nm$^2$). In the upper right corner a vortex is pinned at a step edge. (**d**) Differential conductance spectrum in the centre of the vortex core, showing the spectroscopic signature of a vortex bound state. (**e**) d$I$/d$V$ spectral map as a function of sample bias and distance from the centre of the vortex core along the green line in panel **c** showing the spatial decay of the zero-bias peak shown in **d**.

In summary, these first temperature- and field-dependent tunnelling spectra on a noncentrosymmetric superconductor show that BiPd is well described with a single gap, an isotropic s-wave order parameter and the predictions of BCS theory, confirmed by bulk measurements. Our results suggest that the coupling mechanism that mediates Cooper pairing in BiPd favours s-wave superconductivity, leading to a negligible triplet component[29]. While the superconducting phase of BiPd appears to be topologically trivial, its normal-state band structure exhibits nontrivial features. By means of electronic structure calculations, we have shown that the large atomic SOC of Bi alters the band topology near the Γ point, leading to the appearance of a Dirac-cone surface state, the existence of which is confirmed by our STM measurements of the surface density of states. Our results highlight the importance of a characterization both at the macroscopic and microscopic scales to establish a complete picture of the physical properties, while the ability to apply STM to noncentrosymmetric superconductors will open new avenues of exploration in this exciting field.

## Methods

**Crystal growth.** Large single crystals of BiPd were grown by a modified Bridgman–Stockbarger technique. Stoichiometric quantities of Bi (Alfa Aesar, 99.999%) and Pd (Degussa, 99.95%) were sealed under vacuum in a quartz ampoule having a conical end, then suspended in a vertical tube furnace. While maintaining a temperature gradient over the length of the quartz tube, the furnace was heated well above the melting point to ensure homogeneity. BiPd melts congruently at 600 °C. The ampoule was subsequently cooled through the melting temperature at a rate of 0.2 °C h$^{-1}$, then cooled through the α–β phase transition at 189 °C at 0.5 °C h$^{-1}$ to maximize the domain size of the low-temperature α phase. The resulting crystal readily cleaves perpendicular to the unique monoclinic $b$ axis. Analysis of the composition by energy dispersive X-ray spectroscopy suggests a bismuth deficiency of 1–2%, comparable to the technique's uncertainty. Additional characterization is presented in Supplementary Fig. 4 and Supplementary Note 2.

**Electronic structure calculations.** The electronic structure of BiPd was obtained by a fully relativistic linear muffin-tin orbital calculation[30,31] using the experimental crystal structure of ref. 13. Relativistic effects, including spin–orbit interactions, were fully taken into account by solving the four-component Dirac equation inside each atomic sphere. The latter is crucial to obtain accurate results for the $p_{1/2}$ bands of heavy elements such as Bi[32]. Figure 2c,d show a sketch of the electronic structure, the full calculations are shown in Supplementary Figs 2 and 3 and Supplementary Note 1.

All band structure calculations were performed using the in-house PY LMTO computer code as described in ref. 31. The code is available on demand.

**STM measurements.** All experiments were performed using a home-built STM, operating in cryogenic vacuum at temperatures down to 10 mK and in magnetic fields up to 14 T (ref. 21). Single-crystal samples of BiPd were cleaved *in situ* at low temperatures. We used STM tips cut from a platinum–iridium wire. Bias voltages were applied to the sample with the tip at virtual ground. Differential conductance spectra were recorded through a standard lock-in technique with frequency $f = 433$ Hz.

## Acknowledgements

We thank C. Ast and H. Benia for useful discussions, C. Stahl and E. Goering for help with magnetization measurements, and the Crystal Growth and Chemical Service Groups at MPI-FKF for assistance. Z.S. acknowledges the financial support from Rubicon grant No. 680.50.1119 (NWO, NL), P.W. and A.M. from EPSRC. P.W. acknowledges the support from Max-Planck-Society, DCP from IBS-R009-G1.

## Author contributions

D.C.P. grew samples and performed magnetization, specific heat and transport measurements, E.A.Y. and C.L. performed the transport measurements at low currents, Z.S., M.E. and A.M. performed the STM measurements, Z.S., M.E., A.M., A.P.S., and P.W. analysed data, A.Y. performed the DFT calculations, P.W., A.P.S. and D.C.P. wrote the manuscript. All authors discussed the manuscript.

## Additional information

**Supplementary Information** accompanies this paper at http://www.nature.com/naturecommunications

**Competing financial interests**: The authors declare no competing financial interests.

**Reprints and permission** information is available online at http://npg.nature.com/reprintsandpermissions.

**How to cite this article**: Sun, Z. *et al.* Dirac surface states and nature of superconductivity in Noncentrosymmetric BiPd. *Nat. Commun.* 6:6633 doi: 10.1038/ncomms7633 (2015).